\documentstyle[twoside,fleqn,espcrc2,epsfig]{article}

\def\rmO{{\rm O}}

\def\defeq{\mathrel{\mathop=^{\rm def}}}
\def\proof{\noindent{\sl Proof:}\kern0.6em}

\def\frac#1#2{\hbox{$#1\over#2$}}
\def\dual{\mathstrut^*\kern-0.1em}

\def\lvec#1{\setbox0=\hbox{$#1$}
    \setbox1=\hbox{$\scriptstyle\leftarrow$}
    #1\kern-\wd0\smash{
    \raise\ht0\hbox{$\raise1pt\hbox{$\scriptstyle\leftarrow$}$}}
    \kern-\wd1\kern\wd0}
\def\rvec#1{\setbox0=\hbox{$#1$}
    \setbox1=\hbox{$\scriptstyle\rightarrow$}
    #1\kern-\wd0\smash{
    \raise\ht0\hbox{$\raise1pt\hbox{$\scriptstyle\rightarrow$}$}}
    \kern-\wd1\kern\wd0}

\def\nabstar#1{\nabla\kern-0.5pt\smash{\raise 4.5pt\hbox{$\ast$}}
               \kern-4.5pt_{#1}}

\def\drvstar#1{\partial\kern-0.5pt\smash{\raise 4.5pt\hbox{$\ast$}}
               \kern-5.0pt_{#1}}

\def\momp#1#2{
    \setbox0=\hbox{${#1}$}\setbox1=\hbox{${#1}_{#2}$}
    {#1}_{#2}\kern-\wd1\kern\wd0
    \smash{\raise4.5pt\hbox{$\scriptscriptstyle +$}}}
\def\momm#1#2{
    \setbox0=\hbox{${#1}$}\setbox1=\hbox{${#1}_{#2}$}
    {#1}_{#2}\kern-\wd1\kern\wd0
    \smash{\raise4.5pt\hbox{$\scriptscriptstyle -$}}}
\def\mompm#1#2{
    \setbox0=\hbox{${#1}$}\setbox1=\hbox{${#1}_{#2}$}
    {#1}_{#2}\kern-\wd1\kern\wd0
    \smash{\raise4.5pt\hbox{$\scriptscriptstyle \pm$}}}
\def\smomp#1#2{
    \setbox0=\hbox{${#1}$}\setbox1=\hbox{${#1}_{#2}$}
    {#1}_{#2}\kern-\wd1\kern\wd0
    \smash{\raise3pt\hbox{$\scriptscriptstyle +$}}}
\def\smomm#1#2{
    \setbox0=\hbox{${#1}$}\setbox1=\hbox{${#1}_{#2}$}
    {#1}_{#2}\kern-\wd1\kern\wd0
    \smash{\raise3pt\hbox{$\scriptscriptstyle -$}}}
\def\smompm#1#2{
    \setbox0=\hbox{${#1}$}\setbox1=\hbox{${#1}_{#2}$}
    {#1}_{#2}\kern-\wd1\kern\wd0
    \smash{\raise3pt\hbox{$\scriptscriptstyle \pm$}}}

\def\si{\kern1pt{\rm si}}
\def\co{\kern1pt{\rm co}}

\def\rhoprime{\rho\kern1pt'}
\def\rhobar{\bar{\rho}}
\def\rhobarprime{\rhobar\kern1pt'}
\def\rhobartilde{\kern2pt\tilde{\kern-2pt\rhobar}}
\def\rhobartildeprime{\kern2pt\tilde{\kern-2pt\rhobar}\kern1pt'}

\def\zetabar{\bar{\zeta}}
\def\zetaprime{\zeta\kern1pt'}
\def\zetabarprime{\zetabar\kern1pt'}
\def\zetar{\zeta_{\raise-1pt\hbox{\sixrm R}}}
\def\zetabarr{\zetabar_{\raise-1pt\hbox{\sixrm R}}}

\def\phiimpr{\phi_{\kern0.5pt\hbox{\sixrm I}}}

\def\diracstar#1#2{
    \setbox0=\hbox{$\gamma$}\setbox1=\hbox{$\gamma_{#1}$}
    \gamma_{#1}\kern-\wd1\kern\wd0
    \smash{\raise4.5pt\hbox{$\scriptstyle#2$}}}

\def\ba{b_{\rm A}}
\def\tba{\tilde{b}_{\rm A}}
\def\bp{b_{\rm P}}

\def\bv{b_{\rm V}}
\def\tbv{\tilde{b}_{\rm V}}

\def\bg{b_{\rm g}}
\def\bm{b_{\rm m}}
\def\tbm{\tilde{b}_{\rm m}}
\def\bmu{b_{\mu}}

\def\ca{c_{\rm A}}
\def\cv{c_{\rm V}}

\def\csw{c_{\rm sw}}

\def\f1{f_1}

\def\h1{h_1}

\def\opprime#1{\setbox0=\hbox{${\cal O}$}\setbox1=\hbox{${\cal O}_{\rm #1}$}
    {\cal O}_{\rm #1}\kern-\wd1\kern\wd0
    \smash{\raise4.5pt\hbox{\kern1pt$\scriptstyle\prime$}}\kern1pt}

\def\ophatprime#1{\setbox0=\hbox{$\widehat{\cal O}$}
    \setbox1=\hbox{$\widehat{\cal O}_{\rm #1}$}
    \widehat{\cal O}_{\rm #1}\kern-\wd1\kern\wd0
    \smash{\raise4.5pt\hbox{\kern1pt$\scriptstyle\prime$}}\kern1pt}

\def\bopprime#1{\setbox0=\hbox{${\cal O}$}\setbox1=\hbox{${\cal O}_{\rm #1}$}
    {\cal L}_{\rm #1}\kern-\wd1\kern\wd0
    \smash{\raise4.5pt\hbox{\kern1pt$\scriptstyle\prime$}}\kern1pt}

\def\blagprime#1{\setbox0=\hbox{${\cal B}$}\setbox1=\hbox{${\cal B}_{#1}$}
    {\cal B}_{#1}\kern-\wd1\kern\wd0
    \smash{\raise5.2pt\hbox{\kern1pt$\scriptstyle\prime$}}\kern1pt}

\def\muq{\mu_{\rm q}}
\def\mq{m_{\rm q}}
\def\mqtilde{\widetilde{m}_{\rm q}}
\def\muqtilde{\widetilde{\mu}_{\rm q}}

\def\mc{m_{\rm c}}

\def\za{Z_{\rm A}}
\def\zv{Z_{\rm V}}
\def\zp{Z_{\rm P}}

\def\zg{Z_{\rm g}}
\def\zm{Z_{\rm m}}

\def\Za{\za}
\def\Zv{\zv}
\def\Zp{\zp}

\def\Zg{\zg}
\def\Zm{\zm}

\def\gtilde{\tilde{g}_0}

\def\msbar{{\rm \overline{MS\kern-0.05em}\kern0.05em}}

\begin{document}
\title{
       \vspace{-4.0cm}
       \rightline{\normalsize CERN-TH/2001-279}
       \rightline{\normalsize October 2001}
       \vspace{2.0cm}
Some remarks on O($a$) improved twisted 
mass QCD\thanks{
presented by S.~Sint at the ${\rm XIX}$ 
International Symposium on Lattice Field Theory ``Lattice 2001'', 
August 19 -- 24, 2001, Berlin, Germany}}
\author{
Roberto Frezzotti\address{Universit\`a di Milano-Bicocca, 
Dipartimento di Fisica, Piazza della Scienza 3, I-10126 Milan, Italy}
and Stefan Sint\address{CERN, Theory Division, CH-1211 Geneva 23, Switzerland}
}
\begin{abstract}
Twisted mass QCD (tmQCD) has been introduced as a solution 
to the problem of unphysical fermion zero modes in lattice QCD
with quarks of the Wilson type. We here argue
that O($a$) improvement of the tmQCD action and simple quark bilinear
operators can be more economical than in the standard framework.
In particular, an improved and renormalized estimator of the
pion decay constant in two-flavour QCD is available,
given only the Sheikholeslami-Wohlert coefficient $\csw$ 
and an estimate of the critical mass $\mc$.
\end{abstract}
\maketitle

\section{INTRODUCTION}

Twisted mass QCD (tmQCD) \cite{Frezzotti:2000vv,Frezzotti:2001nk} is a 
theoretically sound method to eliminate unphysical
fermionic zero modes, which are at the origin of
both conceptual and technical problems 
in lattice QCD with quarks of the Wilson type.
In the continuum limit, tmQCD is equivalent 
to QCD with a standard mass term, provided the
parameters and correlation functions are correctly 
matched~\cite{Frezzotti:2001nk}. This physical equivalence
implies that the angle $\alpha$, defined by the
ratio of twisted to standard mass parameter,
\begin{equation}
  \tan(\alpha) = \mu_{\rm R}/m_{\rm R},
\end{equation}
is unphysical.  Furthermore, the matching of theories 
defined at different values of $\alpha$
induces a mapping between composite fields,
which can be used to circumvent certain lattice 
renormalization problems. Examples for this
are the definition of the order parameter of chiral symmetry, 
the pion decay constant, and some matrix elements
of the effective weak hamiltonian~\cite{Frezzotti:2001nk,Guagnelli:2001a}.

Here we want to review some features of O($a$) improved 
tmQCD~\cite{Frezzotti:2001ea,DellaMorte:2001ys}. 
We restrict attention to the action and 
quark bilinear operators which appear in the PCAC and PCVC relations, 
i.e.~the simplest Ward identities associated with chiral and
flavour symmetries.

\section{SET-UP}

The lattice Dirac operator of two-flavour QCD with a chirally twisted
mass term is given by
\begin{equation}
  D_{\rm twist}\defeq D_{\rm W} + m_0+ i\muq\gamma_5\tau^3.
\end{equation}
where $D_{\rm W}$ denotes the massless Wilson-Dirac operator and the Pauli
matrix $\tau^3$ acts in flavour space.
Assuming that O($a$) improvement has been implemented 
in the massless theory~\cite{Luscher:1996sc},
the O($a$) improved bare parameters of the action are given by
\begin{eqnarray}
  \gtilde^2   &=& g_0^2(1+\bg a\mq),\\
  \mqtilde    &=& \mq + b_{\rm m} a\mq^2 +\tbm a\muq^2,\\
  \muqtilde   &=& \muq(1+\bmu a\mq),
\end{eqnarray}
where $\mq=m_0-\mc$ is the subtracted bare mass.
In a mass-independent scheme the renormalized parameters then
take the form
\begin{eqnarray}
  g_{\rm R}^2 &=&  \gtilde^2\Zg(\gtilde^2,a\mu),\\
  m_{\rm R}   &=&  \mqtilde\Zm(\gtilde^2,a\mu),\label{mr}\\
  \mu_{\rm R} &=&  \muqtilde Z_\mu(\gtilde^2,a\mu).
\end{eqnarray}   
O($a$) improvement of the action thus 
introduces the improvement coefficients $b_\mu$ and $\tbm$,
in addition to the standard coefficients $\bm$ and $\bg$.
At this point we note that none of the coefficients proportional
to $\mq$ is needed if $\mq=\rmO(a)$, as their contribution
to physical quantities is then of O($a^2$) 
and thus negligible in the spirit of O($a$) improvement.   
Furthermore, eq.~(\ref{mr}) then implies $m_{\rm R}=\rmO(a)$ so that  
the physical quark mass $(m_{\rm R}^2+\mu_{\rm R}^2)^{1/2}$
is essentially defined by $\mu_{\rm R}$, and~the
angle $\alpha$ is close to $\pi/2$.

We now consider the O($a$) improved bare composite 
fields~\cite{Luscher:1996sc,Frezzotti:2001ea}
\begin{eqnarray}
  (A_{\rm I})_\mu^a &=& A_\mu^a+\ca a\tilde\partial_\mu
                         P^a + a\muq\tba \varepsilon^{3ab} V_\mu^b, \\
  (V_{\rm I})_\mu^a &=& V_\mu^a+\cv a\tilde\partial_\nu
                 T^a_{\mu\nu}+ a\muq\tbv \varepsilon^{3ab} A_\mu^b,\\
  (P_{\rm I})^a &=& P^a,
\end{eqnarray}
with isospin index $a=1,2$. Renormalized improved operators 
are then multiplicatively related to the improved bare ones,
\begin{eqnarray}
  (A_{\rm R})_\mu^a &=& \Za(\gtilde^2)(1+\ba a\mq)(A_{\rm I})_\mu^a, \\
  (V_{\rm R})_\mu^a &=& \Zv(\gtilde^2)(1+\bv a\mq)(V_{\rm I})_\mu^a, \\
  (P_{\rm R})^a     &=& \Zp(\gtilde^2,a\mu)(1+\bp a\mq)(P_{\rm I})^a.
\end{eqnarray}
While the improvement coefficients may be considered functions
of the bare coupling $g_0$, consistent O($a$) improvement 
implies that the $Z$-factors are functions of the improved 
bare coupling $\gtilde^2$~\cite{Luscher:1996sc}. As non-perturbative
determinations of $\bg$ are not available 
(cf., however, \cite{Martinelli:1997zc}), this renders
chiral extrapolations of O($a$) improved matrix elements 
difficult\footnote{This problem does not occur in
the quenched approximation where $\bg=0$.}.
Compared to the standard framework at $\muq=0$ we note
that in O($a$) improved tmQCD 
there are only two new coefficients ($\tbv$ and $\tba$) 
required to improve the above operators. 
Considering again $\alpha=\pi/2$ this means that the massive 
theory is improved with a single new
coefficient $\tbm$ in the action 
(rather than two, $\bg$ and $b_{\rm m}$), 
while the operators are improved with $\tba$ and $\tbv$ as compared
to $\bv,\ba,\bp$. This comparison becomes even more favourable
if one takes into account a generic

\section{REDUNDANCY OF IMPROVEMENT COEFFICIENTS}

To illustrate this point we consider the renormalized 2-point functions
\begin{eqnarray}
   G_{\rm A}(x-y) &=& \left\langle (A_{\rm R})^1_0(x)(P_{\rm R})^1(y)
   \right\rangle,\\
   G_{\rm V}(x-y) &=& \left\langle (V_{\rm R})^2_0(x)(P_{\rm R})^1(y)
  \right\rangle,
\end{eqnarray}
which, for the proper choice of the improvement coefficients,
are expected to converge to their continuum limit with O($a^2$)
corrections. If all improvement coefficients were necessary
one would expect uncancelled O($a$) effects to arise if any
of the improvement coefficients is modified by terms of O(1). 
Denoting such shifts to $\tbm,b_\mu,\tba,\tbv$ by 
$\Delta\tbm$ etc., we find that the induced
O($a$) effect in $G_{\rm A}$ is of the form
\begin{eqnarray}
  \Delta G_{\rm A}(x) \propto a\mu_{\rm R}\Bigl[\Delta \tbm\mu_{\rm R} 
  {\partial\over{\partial m_{\rm R}}}G_{\rm A}(x) \nonumber\\
   \hphantom{a}
   +\Delta b_\mu c_1 m_{\rm R} 
 {\partial\over{\partial\mu_{\rm R}}}G_{\rm A}(x)  
   \hphantom{a}
  + \Delta \tba c_2 G_{\rm V}(x)\Bigr],
 \label{vary}
\end{eqnarray}
with some constants $c_1,c_2$ which can be easily worked 
out~\cite{Frezzotti:2001ea}.
Now, due to the identity in the continuum limit,
\begin{equation}
   \left(m_{\rm R}{\partial\over{\partial\mu_{\rm R}}}
      -\mu_{\rm R}{\partial\over{\partial m_{\rm R}}}\right)G_{\rm A}(x)
  = - G_{\rm V}(x),
 \label{alpha_derivative}
\end{equation}
one concludes that $\Delta\tbm$, $\Delta b_\mu$ and $\Delta \tba$
need not vanish separately for $G_{\rm A}(x)$ to remain O($a$) improved.
We find this to be a generic feature of tmQCD, which can be 
traced back to the equivalence of correlation functions of 
tmQCD and QCD in the continuum limit. 
In fact, the l.h.s.~of eq.(\ref{alpha_derivative}) is nothing
else but the derivative with respect to the unphysical
parameter $\alpha$. 

O($a$) improved tmQCD as introduced above may hence be regarded
as a one-parameter family of O($a$) improved theories. Choosing
$\tbm$ as the free parameter, we set 
\begin{equation}
 \tbm=-\frac12,
 \label{exact}
\end{equation}
{\em exactly}.
The other coefficients are then fixed, and
in perturbation theory given by [$C_{\rm F}=(N^2-1)/2N$]
\begin{eqnarray}
  b_\mu &=& -0.103(3) \, C_{\rm F} g_0^2 +\rmO(g_0^4),\\  
  \tba &=&  \hphantom{+}0.086(4) \, C_{\rm F} g_0^2 +\rmO(g_0^4),\\  
  \tbv &=&  \hphantom{+}0.074(3) \,C_{\rm F} g_0^2 +\rmO(g_0^4).
\end{eqnarray}
Indeed, the choice~(\ref{exact}) is partially motivated by 
the fact that the tree level values of these coefficients then vanish.
Beyond perturbation theory,
$\tba$ can be determined through the PCAC relation~\cite{Guagnelli:2001b},
$\tbv$ can be obtained by imposing the physical parity symmetry 
in tmQCD at finite $a$, whereas the PCVC relation involves
the combination $\bp+b_\mu$.
We also note that tmQCD offers new ways to determine some
of the standard coefficients such as $\ba$, by matching
appropriate correlation functions of tmQCD and standard QCD.
This is not too surprising if one recalls that tmQCD and QCD
are, in the continuum, related by a chiral symmetry transformation,
and given the fact that most of the standard coefficients are determined
by chiral Ward identities~\cite{Bhattacharya:1999uq}.

\section{O($a$) IMPROVED  $F_\pi$}

In tmQCD the flavour symmetry is only softly broken by
the twisted mass term. As a consequence, there exists a vector
current $\tilde{V}_\mu^a$, which satisfies the PCVC relation exactly,
\begin{equation}
   \partial^\ast_\mu\tilde{V}^a_\mu = -2\muq \varepsilon^{3ab}P^b.
\end{equation}
This vector current is protected against renormalization,
which implies $Z_\mu=Z_{\rm P}^{-1}$ in any scheme which
respects the Ward identities. 
Recalling that the vector current in tmQCD at $m_{\rm R}=0$ 
is interpreted as the physical axial current, the pion 
decay constant $F_\pi$ can be extracted from the asymptotic
behaviour of the correlation function,
\begin{equation}
  \muq \tilde{G}_{\rm P}(x_0) = a^3\sum_{\bf x} \muq\left\langle
  P^1(x)(P_{\rm R})^1(0)\right\rangle,
\end{equation}
for large times $x_0$, after division by $m_\pi^2$ and
by the wave function renormalization of the interpolating field $P^1$.
Given the renormalization properties of both $P^1$ and $\muq$ we
conclude that $F_\pi$ is obtained with O($a^2$) errors only.
First studies of $F_\pi$ using this method in the quenched
approximation have been presented in~\cite{DellaMorte:2001}.

Generalizing this procedure to non-vanishing $m_{\rm R}$,
one considers the r.h.s.~of the physical PCAC relation,
\begin{equation}
  \sqrt{m_{\rm R}^2+\mu_{\rm R}^2} \left\langle
  P_{\rm R}^1(x)P_{\rm R}^1(0)\right\rangle.
\end{equation}
From this more general expression one 
infers that $F_\pi$ remains O($a$) improved
even if $m_{\rm R}=\rmO(a)$ rather than O($a^2$)~\cite{Michele}.

\section{CONCLUSIONS}

We have argued that O($a$) improvement of 
tmQCD at $m_{\rm R}=0 \Leftrightarrow \alpha=\pi/2$ 
is more economical than in the standard theory.
No additional free parameters arise in the action, so that
$m_{\rm R}=0$ can be satisfied up to O($a^2$), provided the
coefficients $\csw,\ca$ are known\footnote{The knowledge of
$\ca$ is required if one needs
an O($a$) improved estimate of the critical mass from the PCAC
relation.}. Moreover, an O($a$) improved estimate of 
$F_\pi$ can even be obtained with the knowledge of $\csw$ alone.
This is to be confronted with the standard situation where
also $Z_{\rm A},\bg,c_{\rm A},b_{\rm A}$ are required.
An interesting aspect  of O($a$) improved tmQCD 
is the absence of a quark mass dependent 
rescaling of $g_0$, which allows for chiral extrapolations
to be done at fixed $g_0$, whilst maintaining O($a$) improvement.

\vskip 1ex

This work is part of the ALPHA collaboration research programme.
We thank Peter Weisz for an enjoyable collaboration and critical
comments on a draft of this writeup.
S.~Sint is grateful to the Theory Division of DESY-Hamburg,
where part of this work has been done.

\end{document}